\documentclass{article}

\usepackage{amsmath,amsthm,amssymb,fullpage,hyperref,graphicx}
\usepackage[caption=false]{subfig}
\usepackage{authblk}
\usepackage{xcolor}

\title{Comment on 'Winding around non-Hermitian singularities' by Zhong {\it et al.}, Nat.~Commun.~9, 4808 (2018)}

\author[1,2]{Eric J. Pap}
\author[1]{Dani{\"e}l Boer}
\author[2]{Holger Waalkens}

\affil[1]{Van Swinderen Institute for Particle Physics and Gravity, University of Groningen, Nijenborgh 4, 9747 AG Groningen, The Netherlands}
\affil[2]{Bernoulli Institute, University of Groningen, Nijenborgh 9, NL-9747 AG Groningen, The Netherlands}

\date{\today}

\newcommand{\R}{\mathbb{R}}
\newcommand{\C}{\mathbb{C}}

\begin{document}

\maketitle

In a recent paper by Zhong \textit{et al.} \cite{Zhong2018WindingSingularities} a formalism is presented with which one can calculate the permutations of eigenstates that arise upon encircling (multiple) exceptional points (EPs) in the complex parameter plane of an analytic non-Hermitian Hamiltonian. The authors suggest that upon encircling EPs one should track the eigenvalue branch cuts (BCs) that are traversed, and multiply the associated permutation matrices accordingly. The authors show that the base point and orientation of the traversed loop matter for the outcome. They argue that homotopic loops give the same result and that this depends on the base point considered. In other words, it is not free homotopy that matters, but homotopy of based oriented loops. These insights are fully compatible with the results obtained using a more general method that has been published earlier by us in Ref.~\cite{Pap2018Non-AbelianPoints}, which deals with fundamental loops. Using that method one can obtain the precise relation between loops for all cases, including those with different base points. We will comment on this in more detail below. However, first we want to point out that the method of Ref.~\cite{Zhong2018WindingSingularities} fails to achieve its goal of providing a general formalism, because it turns out to be vulnerable to errors in both practical and theoretical applications, which do not occur in the formalism we presented in Ref.~\cite{Pap2018Non-AbelianPoints}. 

A key tool in Ref.~\cite{Zhong2018WindingSingularities} is an ordering of eigenvalues which is independent of the system parameters. In general, the eigenvalues of complex analytic systems of one variable are multi-valued complex functions, implying that a labelling of the eigenvalues cannot be globally continuous. In Ref.~\cite{Zhong2018WindingSingularities}, one global choice is made by ordering the eigenvalues based on the largest real part, which produces a partial order on $\C\cong \R^2$. However, this needs to be extended to a total order on $\C$ to remove ambiguities. A straightforward solution is to use the lexicographic or dictionary order on $\R^2=\R\times \R$, which extends the scheme by sorting complex numbers with equal real part based on the largest imaginary part.
Without this extension one cannot proceed if for some system parameters the operator has two eigenvalues with equal real part. Although this situation may be rare in applications, certainly it matters for a general homotopy argument. Indeed, such parameter values would form as much an obstruction as the branch points, yet this is due to an external choice instead of an intrinsic property of the system related to the EPs.

Nevertheless, even after extending to a total order the formalism may produce errors. These may occur when the order of the eigenvalues is affected by both the branch cut and the sorting scheme at the same time. To illustrate the latter effect, consider the example system given by the following Hamiltonian depending on the complex variable $z$,
\begin{equation}
    H(z)=\begin{pmatrix}z&0\\0&-z\end{pmatrix}
\end{equation}
which has eigenvalues $\lambda_\pm(z)=\pm z$. The eigenvalue $\lambda_+(z)=z$ changes label when moving the parameter $z$ from the right half-plane to the left half-plane, yet no branch cut is involved. This problem could be mediated by changing the sorting scheme to first check the imaginary part, but then the issue appears with the upper and lower half-planes. At this point one may argue that any total order on $\C$ has such issues, which is due to the fact that the metric topology on $\C\cong \R^2$ cannot be an order topology. 

\begin{figure*}
    \subfloat[\label{fig:path in N}]{\includegraphics{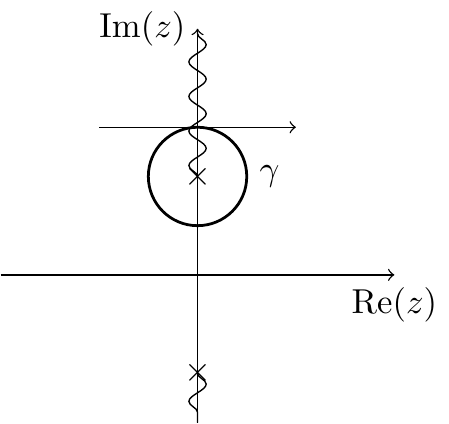}
    }
    \hfill
    \subfloat[\label{fig:sorting issue}
    ]{\includegraphics[width=0.6\textwidth]{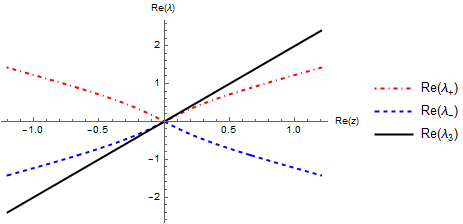}}

    \subfloat[\label{fig:sheets}
    ]{\includegraphics[width=0.5\textwidth]{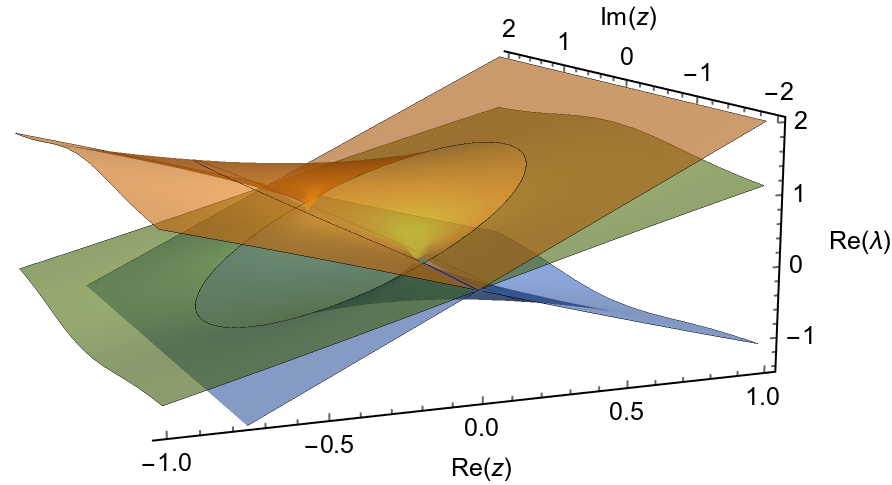}}
    \subfloat[\label{fig:sheets2}
    ]{\includegraphics[width=0.5\textwidth]{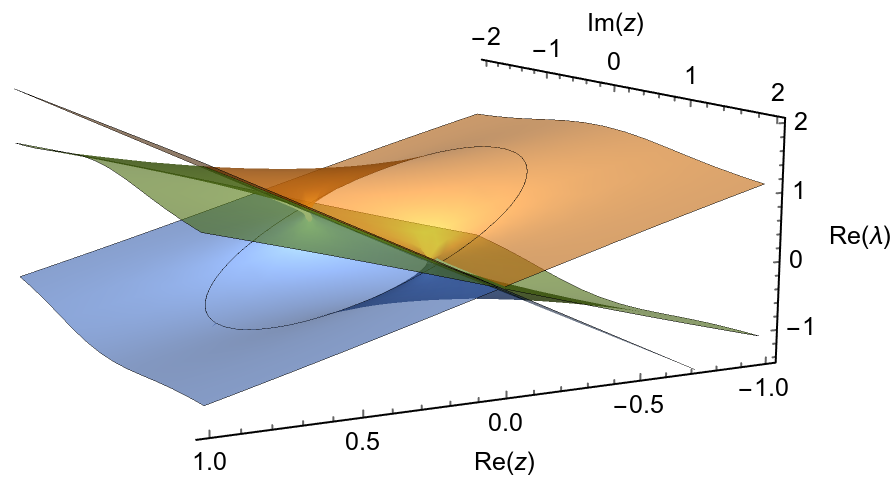}}
    
    \caption{
    Example of a situation where a sorting scheme affects the obtained permutation. In \textbf{a} the situation in parameter space is sketched schematically, consisting of the complex plane with branch cuts, a test loop $\gamma$, and a small arrow marking the path used to find the permutations arising from the crossing of the BC.
    In \textbf{b}, the real part of the eigenvalues on this small path is plotted, parametrized by the real part of the parameter $z$. The imaginary part is fixed at $1.2i$.
    In \textbf{c} and \textbf{d}, the relevant part of  the (real part of the) eigenvalue sheets is shown from two different viewpoints. The colors now indicate the order of the eigenvalues via the pattern orange-green-blue, cf.\ Ref.~\cite{Zhong2018WindingSingularities}. The sheets only intersect at an ellipse at the center.
    Looking at \textbf{b}, we observe that the middle curves on the left and right of the $\mathrm{Re}(\lambda)$-axis are part of the same sheet. From this observation we deduce that the label 2 is fixed. From the other two lines one can similarly deduce that a permutation occurs, which is given by $(13)$. The procedure of Ref.~\cite{Zhong2018WindingSingularities} would then claim that upon encircling $\gamma$ the permutation would be $(13)$. This is not in accordance with the actual result; if one would pick a base point on $\gamma$ just to the left of the BC, then the correct permutation can be seen to be $(12)$. Similarly, the correct permutation is given by $(23)$ on the right as there the labelling is different.    }
    \label{fig:wrong perm mat}
\end{figure*}

For each branch cut, its permutation matrix should be found by picking an infinitesimal trajectory across the branch cut and comparing the initial and final positions of the eigenvalues in the sorting scheme. As observed in the example above, the labelling can change without crossing a branch cut. This is relevant only if the change due to the sorting occurs exactly when intersecting the branch cut, as otherwise one can pick another infinitesimal path so that the sorting problem vanishes. We illustrate this by the system
\begin{equation}
    T(z)=\begin{pmatrix} 1 & z & 0\\ z & -1 & 0 \\ 0 & 0 & 2z\end{pmatrix}
\end{equation}
which models an EP2 system with the first $2\times 2$ block, and a separate third level in the $1\times 1$ block which is unrelated to the EP2 topology. This can be seen from the eigenvalues, which are given by
\begin{equation}
    \lambda_\pm(z)=\pm \sqrt{1+z^2} , \quad \lambda_3(z)=2z. 
\end{equation}
The branch points are located at $z_\pm=\pm i$, and we define BCs along the imaginary axis below $z_-$ and above $z_+$. Hence we can restrict our attention to an open strip along the $\mathrm{Im}(z)$-axis. We note that the eigenvalues are all imaginary at the BCs and do not change order on the cuts. Let us now encircle $z_+$ with a small circle; this means that we cross the upper BC once, and no other BCs need to be considered, see Fig.~\ref{fig:path in N}. As the eigenvalues are imaginary on the entire BC, the exact point where the encircling loop and the BC intersect is irrelevant. The formalism of Ref.~\cite{Zhong2018WindingSingularities} prescribes that the obtained interchange of eigenvalues is given by the permutation matrix at the branch cut. As seen in Fig.~\ref{fig:wrong perm mat}, in this way one would conclude that one of the EP eigenvalues is interchanged with the third level, which is clearly false.

One could try to alter the proposed procedure in such a way that errors like that of Fig.~\ref{fig:wrong perm mat} no longer appear. The main cause of the error is the use of a parameter independent ordering of the eigenvalues, hence switching to locally defined labels of the eigenvalues seems to be a straightforward fix. Yet, the method in Ref.~\cite{Zhong2018WindingSingularities} is devised to exploit a global labelling. If the labels are defined only locally, i.e.\ per patch of parameter space, one should also draw the boundaries of these patches and associate permutation matrices to them as well. Although this is possible, the convenience of the method would be reduced significantly.

A different approach is to associate the permutations of eigenvalues not to the BCs but to fundamental loops encircling the singularities \cite{Pap2018Non-AbelianPoints}. In short, the idea is to (homotopically) decompose a loop which encloses multiple EPs into more simple loops, e.g.\ loops encircling one EP only. The space inside which these loops must stay is the subspace $X$ of the parameter space consisting of all parameter values for which the Hamiltonian is non-degenerate. For generic complex analytic systems of one variable, the space $X$ is a punctured complex plane like in Ref.~\cite{Zhong2018WindingSingularities}, but for other systems $X$ may be higher dimensional, e.g.\ 3-dimensional space where exceptional lines have been removed. 
Furthermore, one must fix a base point $x_0\in X$, i.e.\ a reference value for all system parameters, at which to compare the eigenvalues. It is only at this point that a labelling of eigenvalues is used, which may be picked arbitrarily and without a sorting scheme, hence it is free of the errors discussed above. Of course, once a labelling is fixed one should stick to it in order to work consistently. The next step in the approach is to consider the fundamental group $\pi_1(X,x_0)$: the group of $x_0$-based loops in $X$ modulo homotopy fixing the initial and final point. This allows one to decompose any based loop of interest into generating loops of $\pi_1(X,x_0)$, which are known as fundamental loops (FLs). By finding the permutations of each FL and composing these according to the FL decomposition, one finds the permutation of the loop one started with. In this way, the FLs replace the BCs as the objects to which one associates a permutation. For the group theoretical details we refer to Ref.~\cite{Pap2018Non-AbelianPoints}.

The FL method has several advantages. First, if $x_0$ can be connected to $x_1\in X$ using a continuous path in $X$, it follows that the permutations at $x_0$ equal those at $x_1$ up to a fixed conjugation operation determined by the permutation associated to the connecting path. In this way one can relate results of loops with different base points, making the whole procedure fully unambiguous, in contrast to what was concluded in Ref.~\cite{Ryu2012AnalysisHamiltonian}.
Second, the computation of the FL permutation can be done numerically. For example, one can discretize the loop in question and fix an order of the eigenvalue at the starting point, which is the base point $x_0$. One can then calculate the eigenvalues at the next point, and update the ordered list by replacing an old eigenvalue with the closest new eigenvalue. This method is accurate for a small enough step size by non-degeneracy of the eigenvalues. By continuing this procedure until one returns to the base point, one can read off the permutation in the ordered list.
Third, group theory allows to identify non-homotopic loops which induce the same permutation, referred to as accidental equivalence in Ref.~\cite{Zhong2018WindingSingularities}. This question reduces to computing the subgroup of $\pi_1(X,x_0)$ of classes that induce the identity permutation.
Finally, the method can be applied to more general cases of degeneracies where the multi-valuedness is not necessarily connected to functions of a complex variable and the sheet structure is not necessarily a Riemann sheet structure. One can even show that the method applies in case the sheets are just continuous, for which we have a paper in preparation. 

It may furthermore be interesting to point out that in the example system in our paper, which has a 3D parameter space, certain EPs were identified which exhibit a different order (EP2 versus EP3) depending on how one encircles them in the parameter space. This was recently also discussed in Ref.~\cite{Zhang2018HybridSystem} where such points are called hybrid EPs. As explained in our paper, the example system, modelled by a $3\times 3$ Hamiltonian, could be experimentally investigated using a three wave-guide system (following investigations like in Ref.~\cite{Schnabel2017PTPoint}). In particular, it could be used to establish experimentally the noncommutative (non-Abelian) nature of multiple EPs. Again we refer to Ref.~\cite{Pap2018Non-AbelianPoints} for further details.
 
\bibliographystyle{naturemag}

\begin{thebibliography}{1}
\expandafter\ifx\csname url\endcsname\relax
  \def\url#1{\texttt{#1}}\fi
\expandafter\ifx\csname urlprefix\endcsname\relax\def\urlprefix{URL }\fi
\providecommand{\bibinfo}[2]{#2}
\providecommand{\eprint}[2][]{\url{#2}}

\bibitem{Zhong2018WindingSingularities}
\bibinfo{author}{Zhong, Q.}, \bibinfo{author}{Khajavikhan, M.},
  \bibinfo{author}{Christodoulides, D.~N.} \& \bibinfo{author}{El-Ganainy, R.}
\newblock \bibinfo{title}{{Winding around non-Hermitian singularities}}.
\newblock \emph{\bibinfo{journal}{Nature Communications}}
  \textbf{\bibinfo{volume}{9}}, \bibinfo{pages}{4808} (\bibinfo{year}{2018}).

\bibitem{Pap2018Non-AbelianPoints}
\bibinfo{author}{Pap, E.~J.}, \bibinfo{author}{Boer, D.} \&
  \bibinfo{author}{Waalkens, H.}
\newblock \bibinfo{title}{{Non-Abelian nature of systems with multiple
  exceptional points}}.
\newblock \emph{\bibinfo{journal}{Physical Review A}}
  \textbf{\bibinfo{volume}{98}}, \bibinfo{pages}{023818} (\bibinfo{year}{2018}).

\bibitem{Ryu2012AnalysisHamiltonian}
\bibinfo{author}{Ryu, J.~W.}, \bibinfo{author}{Lee, S.~Y.} \&
  \bibinfo{author}{Kim, S.~W.}
\newblock \bibinfo{title}{{Analysis of multiple exceptional points related to
  three interacting eigenmodes in a non-Hermitian Hamiltonian}}.
\newblock \emph{\bibinfo{journal}{Physical Review A}}  
 \textbf{\bibinfo{volume}{85}}, \bibinfo{pages}{042101} (\bibinfo{year}{2012}).

\bibitem{Zhang2018HybridSystem}
\bibinfo{author}{Zhang, X.-L.} \& \bibinfo{author}{Chan, C.~T.}
\newblock \bibinfo{title}{{Hybrid exceptional point and its dynamical
  encircling in a two-state system}}.
\newblock \emph{\bibinfo{journal}{Physical Review A}} \textbf{\bibinfo{volume}{98}},
  \bibinfo{pages}{33810} (\bibinfo{year}{2018}).

\bibitem{Schnabel2017PTPoint}
\bibinfo{author}{Schnabel, J.}, \bibinfo{author}{Cartarius, H.},
  \bibinfo{author}{Main, J.}, \bibinfo{author}{Wunner, G.} \&
  \bibinfo{author}{Heiss, W.~D.}
\newblock \bibinfo{title}{{PT-symmetric waveguide system with evidence of a
  third-order exceptional point}}.
\newblock \emph{\bibinfo{journal}{Physical Review A}}  
 \textbf{\bibinfo{volume}{95}}, \bibinfo{pages}{053868} 
(\bibinfo{year}{2017}).

\end{thebibliography}

\end{document}